\documentclass[aps,prl,twocolumn,superscriptaddress,floatfix,sort&compress,longnamesfirst]{revtex4-2}

\usepackage{amsmath,amssymb}
\usepackage{bm,bbm}
\usepackage{hyperref}
\usepackage{dcolumn}
\hypersetup{
	colorlinks=true,
	linkcolor=blue,
	filecolor=blue,
	urlcolor=blue,
	citecolor=blue,
}
\usepackage{graphicx}

\begin{document}

\title{Exact steady state of quantum van der Pol oscillator: critical phenomena and enhanced metrology}

\author{Yaohua Li}
\affiliation{State Key Laboratory of Low-Dimensional Quantum Physics, Department of Physics, Tsinghua University, Beijing 100084, P. R. China}
\author{Xuanchen Zhang}
\affiliation{State Key Laboratory of Low-Dimensional Quantum Physics, Department of Physics, Tsinghua University, Beijing 100084, P. R. China}
\author{Yong-Chun Liu}
\email{ycliu@tsinghua.edu.cn}
\affiliation{State Key Laboratory of Low-Dimensional Quantum Physics, Department of Physics, Tsinghua University, Beijing 100084, P. R. China}
\affiliation{Frontier Science Center for Quantum Information, Beijing 100084, China}

\date{\today}

\begin{abstract}
	Quantum criticality of open many-body systems has attracted lots of interest for emergent phenomena and universality. Here we present the exact steady state of the quantum van der Pol oscillator using the complex $P$-representation. We show the threshold corresponds to a dissipative phase transition with abrupt changes of steady-state properties and enhanced metrology. The critical behaviors and finite-size effects are investigated through the analytical steady state. Moreover, we obtain divergent quantum Fisher information (QFI) in the thermodynamic limit both at the critical point and in the time crystal phase, but only the QFI at the critical point approaches the Heisenberg limit. We further prove that the steady-state photon number is the optimized estimated observable with the largest signal-to-noise ratio. We show that the Heisenberg-limited metrology originates from the larger enhancement of the susceptibility than the standard deviation of the photon number. Our work reveals the underlying relation between the time crystal, dissipative phase transition, and enhanced metrology.
\end{abstract}
\maketitle

{\it Introduction.}---Quantum phase transition is a phase transition at zero temperature driven by quantum fluctuations \cite{vojta_quantum_2003}. It is related to the low-energy spectrum. At the phase boundary of a continuous quantum phase transition, the gap between the ground state and the first excited state disappears, and it results in nonanalytical changes of the ground state when crossing the phase boundary. The nonanalytical behaviors near the phase boundary, although induced by quantum effects, can still be understood by the general phase transition theory first developed by Landau. According to the critical exponents that characterize the critical phenomena, the quantum phase transitions can be divided into different universality classes.

Quantum criticality also exists in open systems \cite{hohenberg_theory_1977,landi_nonequilibrium_2022}. For an open system, the Hamiltonian is not enough to describe the system dynamics. The couplings to the environment must be considered. A simple consideration is to assume the environment is a huge and memoryless thermal bath (Markov assumption). Then the system dynamics is governed by a master equation or the Liouville superoperator. Quantum criticality emerges when the gap near the steady state disappears in the thermodynamic limit \cite{kessler_dissipative_2012,minganti_spectral_2018}. Consequently, quantum phase transition in open systems (dissipative phase transition) denotes a phase transition of the steady state.

Dissipative phase transitions have been widely investigated in spin systems \cite{walls_non-equilibrium_1978,werner_phase_2005,diehl_dynamical_2010,marcuzzi_universal_2014,cai_probing_2022,sierant_dissipative_2022,ferioli_non-equilibrium_2023}, optical systems \cite{degiorgio_analogy_1970,marino_driven_2016,casteels_quantum_2017,casteels_critical_2017,fink_signatures_2018,rota_quantum_2019,lieu_symmetry_2020,seibold_dissipative_2020,soriente_distinctive_2021,minganti_liouvillian_2021,minganti_continuous_2021,li_dissipative_2022,bakker_driven-dissipative_2022,minganti_dissipative_2023,bibak_dissipative_2023} and hybrid spin-optical systems such as Dicke \cite{kirton_introduction_2019,nagy_critical_2011,bhaseen_dynamics_2012,zhu_nonreciprocal_2024,brange_lee-yang_2024} and Rabi models \cite{ashhab_superradiance_2013,cai_observation_2021,de_filippis_signatures_2023,zheng_observation_2023,lyu_multicritical_2024}.
A general method to investigate the dissipative phase transition in these systems is the mean-field theory, which neglects finite-size and quantum effects. Beyond the mean-field approximation, the dissipative phase transition can also be analytically solved based on Keldysh field theory \cite{sieberer_dynamical_2013,torre_keldysh_2013,sieberer_keldysh_2016,hwang_dissipative_2018,zhang_driven-dissipative_2024,lin_decoding_2024}, which is a powerful approach leveraging the well-established techniques in equilibrium field theory. Among these open many-body systems, exactly-solvable models are particularly attractive. The exact results not only provide a full description of the phase diagram but also allow us to further investigate the underlying phenomena such as criticality-enhanced metrology \cite{braun_quantum-enhanced_2018,rams_at_2018,garbe_critical_2020,di_candia_critical_2023}. Several exact solutions have been obtained in both spin models \cite{carmichael_analytical_1980,roberts_exact_2023} and optical models \cite{drummond_quantum_1980,bartolo_exact_2016,ben_arosh_quantum_2021} with the emergence of dissipative phase transitions. However, two important aspects of these models: critical phenomena and criticality-enhanced metrology remain unclear. These aspects have been addressed via extensive numerical analysis \cite{zanardi_quantum_2008,raghunandan_high-density_2018} and experimental protocols \cite{ding_enhanced_2022,roques-carmes_biasing_2023,petrovnin_microwave_2024,beaulieu_criticality-enhanced_2024}, while the exact results may shed light on the detailed relation between the quantum criticality and the enhanced metrology.

Here we show the quantum van der Pol oscillator exhibits a dissipative phase transition at the threshold. The phase above the threshold is a time crystal phase. We obtain the exact stationary photon distribution through the Fokker-Planck equations for the complex $P$-representation of the density matrix. We can then obtain the exact results of the Wigner function and photon statistics. The Wigner function exhibits a transition from localized distribution to a quantum limit cycle. Importantly, it allows us to investigate the critical phenomena of the dissipative phase transition, including the critical exponents and the resulting criticality-enhanced metrology. This model has the same finite-size exponent as the open Dicke model and the open Rabi model, but a different photon flux exponent. Moreover, we show that the quantum Fisher information (QFI) $F_{g}$ (with respect to the pump parameter $g$) diverges in the thermodynamic limit both at the critical point and in the time crystal phase. However, only the QFI at the critical point can approach the Heisenberg limit $F_{g}\propto N_{\mathrm{a}}^{2}$, where $N_{a}$ is the steady-state photon number. The QFI in the time crystal phase can only approach the standard quantum limit $F_{g}\propto N_{\mathrm{a}}$. Fortunately, we find the steady-state photon number $\hat{N}_{\mathrm{a}}$ is the optimal observable for the estimated parameter $g$. It allows us to analyze the origin of the Heisenberg-limit sensitivity. When considering the same steady-state photon number, the standard deviation and susceptibility of the photon number at the critical point are both enhanced with double scaling exponents compared with those in the time crystal phase. It results in the sensitivity enhanced from the standard quantum limit $F_{g}\propto N_{\mathrm{a}}$ into the Heisenberg limit $F_{g}\propto N_{\mathrm{a}}^{2}$. Similar scaling can also be obtained for the time required to reach the steady state: $F_{g}\propto T^{2}$ at the critical point and $F_{g}\propto T$ in the time crystal phase.


\begin{figure}[t]
	\centering
	\includegraphics[width=0.48\textwidth]{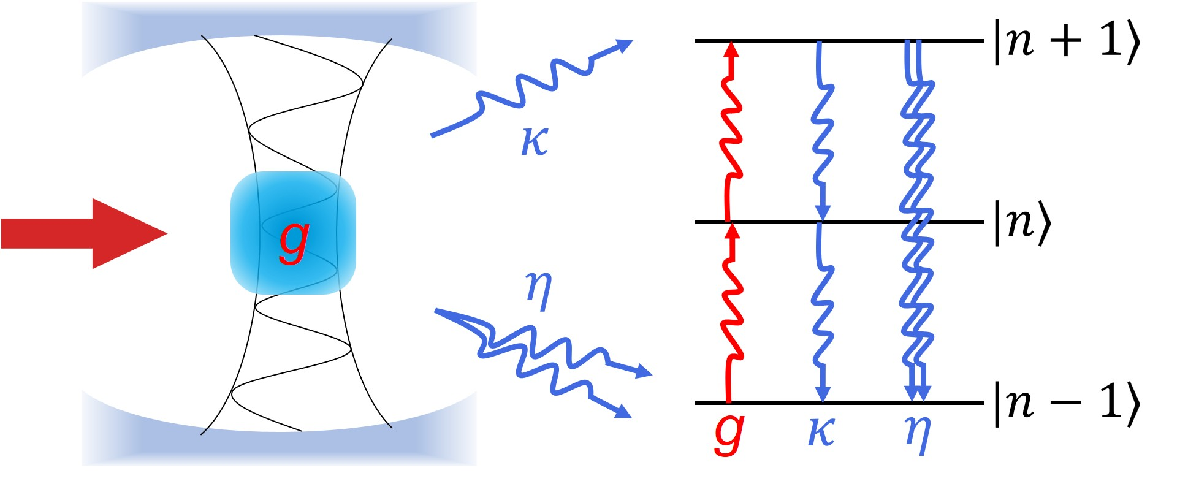}
	\caption{Sketch of our setup: a cavity mode $a$ with single-photon absorption ($g$), single-photon emission ($\kappa$), and two-photon emission ($\eta$). The right panel is a detailed representation of the three processes in the Fock space.}
	\label{fig:a}
\end{figure}

{\it System model.}---We consider a nonlinear bosonic cavity, as shown in Fig. \ref{fig:a}. The cavity is subjected to an incoherent energy pump through one-photon absorption ($g$) and energy dissipation from one-photon emission ($\kappa$) and two-photon emission ($\eta$). This model is also known as the quantum van der Pol oscillator \cite{lee_quantum_2013,eneriz_degree_2019,dutta_critical_2019,wachtler_topological_2023}, and it can be described by the master equation as ($\hbar=1$)
\begin{equation}\label{eq:master}
	\dot{\rho} = i[\rho, \hat{H}] + \kappa \mathcal{D}[\hat{a}]\rho + g \mathcal{D}[\hat{a}^{\dag}]\rho + \eta \mathcal{D}[\hat{a}^{2}]\rho\equiv\hat{\mathcal{L}}\rho,
\end{equation}
with $\hat{H}=\omega_{0}\hat{a}^{\dag}\hat{a}$. Here $\omega_{0}$ is the frequency of the cavity mode, $\hat{a}$ ($\hat{a}^{\dag}$) is the bosonic annihilation (creation) operator, and $\mathcal{D}(\hat{o})\rho=\hat{o}\rho \hat{o}^{\dag}-(\hat{o}^{\dag}\hat{o}\rho+\rho \hat{o}^{\dag}\hat{o})/2$ is the Lindblad dissipator for operator $\hat{o}$. $\hat{\mathcal{L}}$ denotes the Liouville superoperator.

{\it Exact steady state.}---The steady-state density matrix can be obtained by letting $\dot{\rho}=0$. If we are only interested in the steady state, we can introduce a transformation into the rotating frame with frequency $\omega_{0}$. This transformation will not change the steady state, as the steady state does not rotate. Afterwards, we consider the complex $P$-representation of the density matrix \cite{drummond_quantum_1980,bartolo_exact_2016}
\begin{equation}\label{eq:Pre}
	\rho=\int_{\mathcal{C}} d\alpha\int_{\mathcal{C}^{\prime}} d\beta \frac{|\alpha \rangle\langle \beta^{*}|}{\langle \beta^{*}|\alpha \rangle}P(\alpha,\beta),
\end{equation}
where the integral is defined in closed contours that encircle all the singularities. The complex P-representation resolves the limitations of Glauber-Sudarshan P-representation by allowing complex-valued distribution functions. Compared to the Wigner representation which can exhibit negativity, the complex P-representation avoids interpretational ambiguities by using complex integrals. Then the master equation can be transferred into a Fokker-Planck equation as
\begin{equation}
	\partial_{j}(\ln P)=\sum_{i=\alpha,\beta}({\bm{D}}^{-1})_{ji}\left[2A^{i}-\sum_{k=\alpha,\beta}\partial_{k}(D^{ik})\right],
\end{equation}
where
\begin{equation}
	\vec{A} = \begin{pmatrix}
		-i\frac{\kappa-g}{2}\alpha-i\eta\alpha^{2}\beta\\
		-i\frac{\kappa-g}{2}\beta-i\eta\alpha\beta^{2}
	\end{pmatrix},
	\bm{D} = \begin{pmatrix}
		-i\eta\alpha^{2} & ig\\
		ig & -i\eta\beta^{2}
	\end{pmatrix}.
\end{equation}
It gives the complex $P$-representation of the steady-state density matrix as \cite{sup}
\begin{equation}\label{eq:P-steady}
    P(\alpha,\beta) \propto e^{2\alpha\beta}(\eta\alpha\beta-{g})^{\frac{\kappa+g-2\eta}{\eta}}.
\end{equation}
This equation represents the main result of this paper. From the complex $P$-representation we can further obtain the Wigner distribution
\begin{equation}\label{eq:wigner-exact}
    \begin{split}
        W(z)=&\frac{2}{\pi}\int_{\mathcal{C}} d\alpha\int_{\mathcal{C}^{\prime}} d\beta P(\alpha,\beta)e^{-2(\alpha-z)(\beta-z^{*})}\\
        =&\frac{2}{\pi}e^{-2|z|^{2}}\frac{{_{0}F_{1}}(\frac{\kappa+g}{\eta},\frac{4g|z|^{2}}{\eta})}{{_{1}F_{1}}(1;\frac{\kappa+g}{\eta},\frac{2g}{\eta})},
    \end{split}
\end{equation}
and the factorial moments
\begin{equation}\label{eq:nmfp}
    \begin{split}
        \langle \hat{a}^{\dag m}\hat{a}^{m}\rangle=&\int_{\mathcal{C}} d\alpha\int_{\mathcal{C}^{\prime}} d\beta P(\alpha,\beta)\alpha^{m}\beta^{m}\\
        =&\frac{m!\left(\frac{g}{\eta}\right)^{m}{_{1}F_{1}}\left(1+m;\frac{\kappa+g}{\eta}+m,\frac{2g}{\eta}\right)}{\left(\frac{\kappa+g}{\eta}\right)_{m}{_{1}F_{1}}\left(1;\frac{\kappa+g}{\eta},\frac{2g}{\eta}\right)},
    \end{split}
\end{equation}
where $(x)_{n}=x(x+1)\cdots(x+n-1)$ is the Pochhammer symbol, and $_{p}F_{q}$ is the generalized hypergeometric function. Moreover, the density matrix itself can be obtained from Eq. (\ref{eq:Pre}) and Eq. (\ref{eq:P-steady}), which is
\begin{equation}\label{eq:pn}
    \rho=\sum_{n,m}|n\rangle\langle m|\frac{\left(\frac{g}{\eta}\right)^{n}{_{1}F_{1}}\left(1+n;\frac{\kappa+g}{\eta}+n;\frac{g}{\eta}\right)}{\left(\frac{\kappa+g}{\eta}\right)_{n}{_{1}F_{1}}\left(1;\frac{\kappa+g}{\eta};\frac{2g}{\eta}\right)}\delta_{nm}.
\end{equation}
We note Eq. (\ref{eq:nmfp}) and (\ref{eq:pn}) can also be obtained by directly solving the master equation in the Fock basis \cite{hildred_photon_1980,dodonov_exact_1997}. Interestingly, the off-diagonal elements of the steady-state density matrix are all zero, and the information about the steady state is only encoded in the diagonal elements.

\begin{figure}[t]
	\centering
	\includegraphics[width=0.48\textwidth]{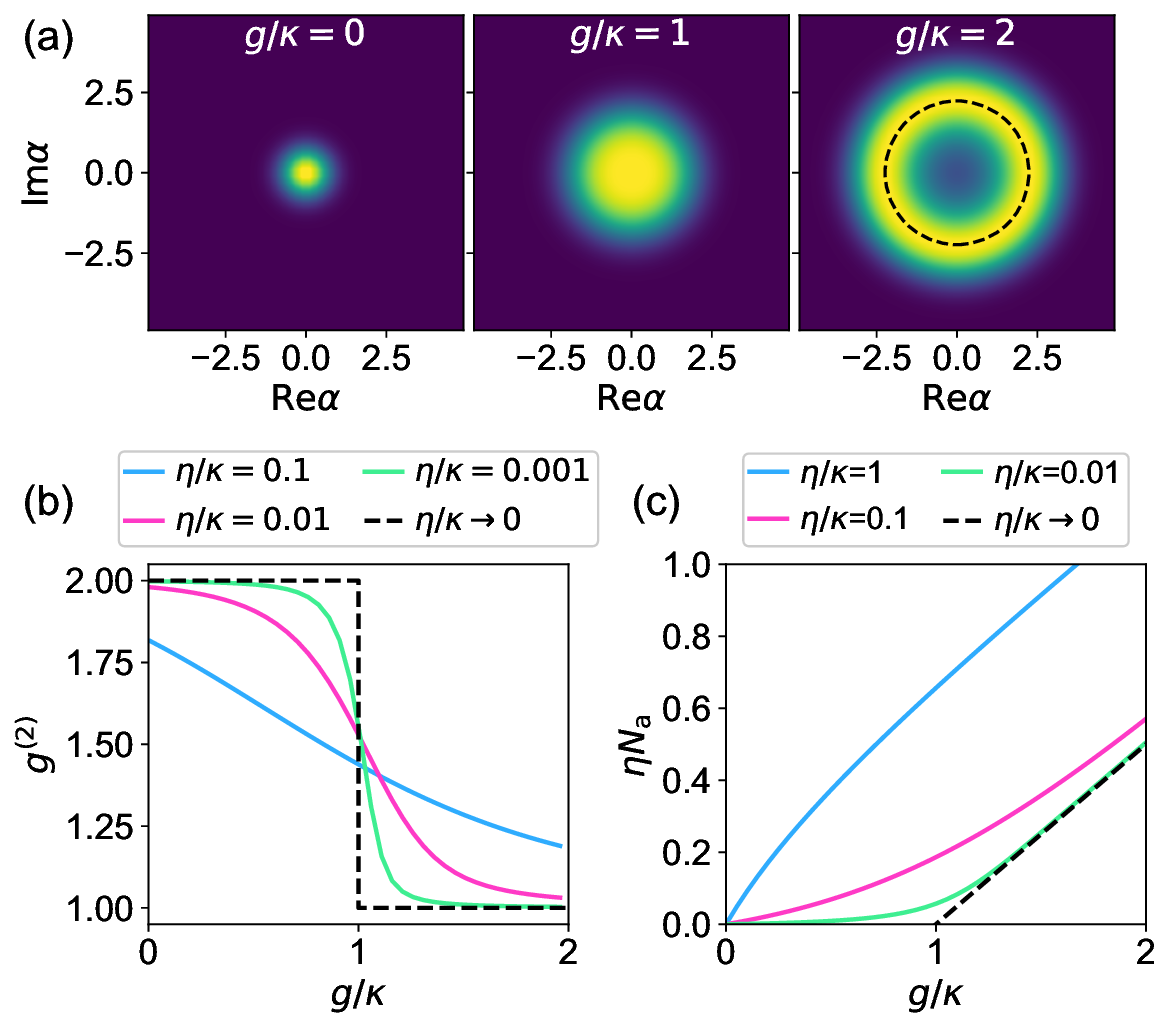}
	\caption{Dissipative phase transition in the steady state. (a) Wigner distributions of the steady state given by Eq. (\ref{eq:wigner-exact}). The nonlinear emission rate is $\eta/\kappa=0.1$. (b),(c) The second-order correlation function $g^{(2)}$ and the rescaled photon number $\eta N_{\mathrm{a}}$ in the steady state.}
	\label{fig:b}
\end{figure}

{\it Dissipative phase transition.}---There are two different phases in this model: the normal phase for $g<\kappa$ and the time crystal phase for $g>\kappa$. The time crystal phase is characterized by the emergence of purely imaginary eigenvalues of the Liouville superoperator or spontaneous stable oscillation in the thermodynamic limit \cite{iemini_boundary_2018,li_time_2024,iemini_floquet_2024,shukla_prethermal_2025}. In the thermodynamic limit, the threshold $g=\kappa$ is a critical point of dissipative phase transition between two phases. Here the thermodynamic limit is defined as $\eta\to0$ \cite{rice_photon_1994,hwang_quantum_2015,carmichael_breakdown_2015,hwang_quantum_2016}. In Fig. \ref{fig:b}(a), we plot the Wigner distribution of the steady state given by Eq. (\ref{eq:wigner-exact}). Although the parameter is still far away from the thermodynamic limit, the two phases exhibit distinct Wigner distributions in the steady state. In the normal phase, the Wigner distribution is localized and takes the maximum value at the origin. In the time crystal phase, the Wigner distribution is uniformly distributed along a circle, as known as the quantum limit cycle \cite{lorch_laser_2014,navarrete-benlloch_general_2017,ben_arosh_quantum_2021}. The radius of the circle can be obtained by employing the mean-field approximation. In this case, the quantum limit cycle reduces to a classical limit cycle, and the radius is given by $r=\sqrt{(g-\kappa)/2\eta}$.

The phase transition is more clear when considering expectation values of physical quantities in the steady state, such as the second-order correlation function $g^{(2)}$ [Fig. \ref{fig:b}(b)] and the rescaled photon number $\eta N_{\mathrm{a}}$ [Fig. \ref{fig:b}(c)]. We can observe a sharp change of the second-order correlation function from 2 to 1 at the critical point when the single-photon absorption rate exceeds the single-photon emission rate. It corresponds to a transition from super-Poisson distribution into Poisson distribution. In the thermodynamic limit, the steady-state photon number remains a finite value for $g<\kappa$ but approaches infinity for $g>\kappa$. Moreover, we can observe that the rescaled photon number $\eta N_{\mathrm{a}}$ exhibits a continuous transition, as shown in Fig. \ref{fig:b}(c). Thanks to the analytical results given by Eq. (\ref{eq:nmfp}), we can analytically obtain the limit of the factorial moments when $\eta \to0$ as \cite{sup}
\begin{equation}\label{eq:aam}
    \lim_{\eta\to0}\langle \hat{a}^{\dag m}\hat{a}^{m}\rangle=\begin{cases}
        \frac{\Gamma(m+1)g^{m}}{(\kappa-g)^{m}}, g<\kappa\\
		\frac{\Gamma\left[(m+1)/2\right]}{\Gamma\left(1/2\right)}\left(\frac{\kappa}{\eta}\right)^{m/2}, g=\kappa\\
        \left(\frac{g-\kappa}{2\eta}\right)^{m}, g>\kappa
    \end{cases}.
\end{equation}
The black dashed lines in Fig. \ref{fig:b}(b) and \ref{fig:b}(c) denotes the analytical results, which match well with the numerical results.

{\it Critical exponents.}---The analytical results of the steady state also allow us to explicitly investigate the critical exponents of the dissipative phase transition. Specifically, we obtain two critical exponents as follows,
\begin{gather}
    \label{eq:na-cri1} \lim_{\eta\to0,\delta_{g} \to 0^{+}}\eta N_{\mathrm{a}} \propto \delta_{g}^{\omega_{1}}, \omega_{1} = 1,\\
    \label{eq:na-cri2} \lim_{\eta\to0} N_{\mathrm{a}}\Bigg|_{\delta_{g} = 0} \propto \eta^{\omega_{2}}, \omega_{2} = -1/2,
\end{gather}
where $\delta_{g}=g-\kappa$ is the distance from the critical point. $\omega_{2}$ is the same as the corresponding exponent in the open Dicke model \cite{sieberer_keldysh_2016} and the open Rabi model \cite{hwang_dissipative_2018}, while the photon flux exponent $\omega_{1}$ is different ($\omega_{1}=-1$ in the latter cases).

\begin{figure}[b]
	\centering
	\includegraphics[width=0.48\textwidth]{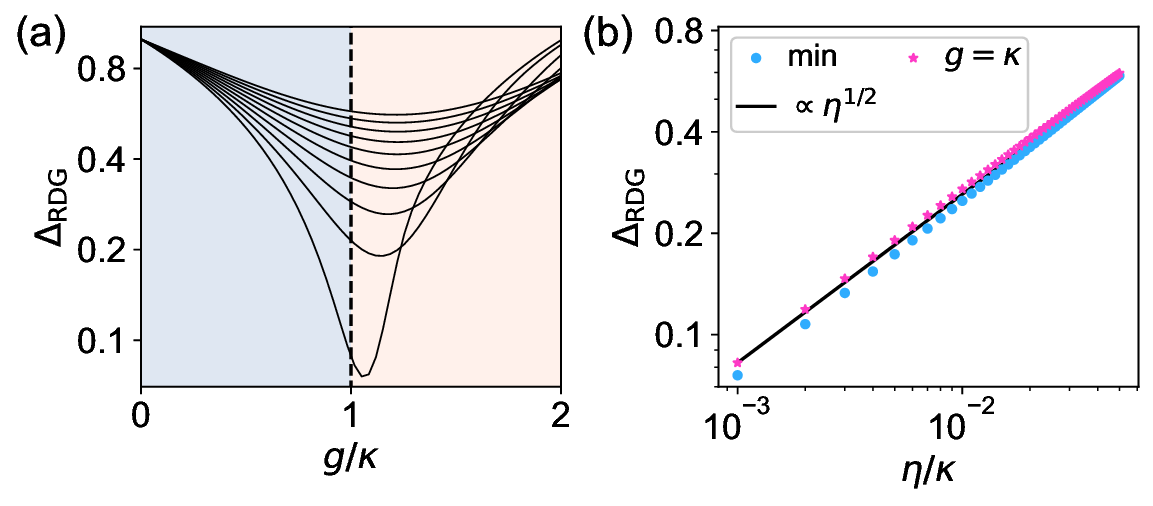}
	\caption{Dissipative gap and critical phenomenon. (a) Dissipative gap as a function of $g$ for different $\eta$. The black dashed line indicates the critical point, and the pink (blue) area indicates the time crystal (normal) phase. The ten lines from the bottom up correspond to equally spaced $\eta/\kappa$ from 0.001 to 0.05. (b) Dissipative gap at the critical point (red dotted line) and at the minimum (blue dotted line). The black line is a reference line corresponding to the critical exponent $1/2$.}
	\label{fig:c}
\end{figure}

As an analog of the quantum phase transition, the quantum criticality in a dissipative phase transition originates from the closing of the Liouvillian spectral gap and the emergence of steady-state degeneracy. In general, the Liouville superoperator is non-Hermitian and has a complex spectrum. Thus, the emergence of steady-state degeneracy requires the closing of the Liouvillian spectral gap from both the real and imaginary axes. In our model, the dynamics of elements in different diagonal lines of the density matrix are independent, and the Liouville superoperator $\hat{\mathcal{L}}$ can be block diagonalized. For the subspace consisted by the $m$th diagonal elements, the imaginary parts of the eigenvalues are always $im\omega_{0}$. Consequently, we can define a dissipative gap $\Delta_{\mathrm{RDG}}$ as the opposite of the largest real nonzero eigenvalue of the Liouville superoperator. The vanishing of this dissipative gap indicates a nonanalytical change of the steady state and thus is a fundamental signature of the dissipative phase transition.

\begin{figure}[t]
	\centering
	\includegraphics[width=0.48\textwidth]{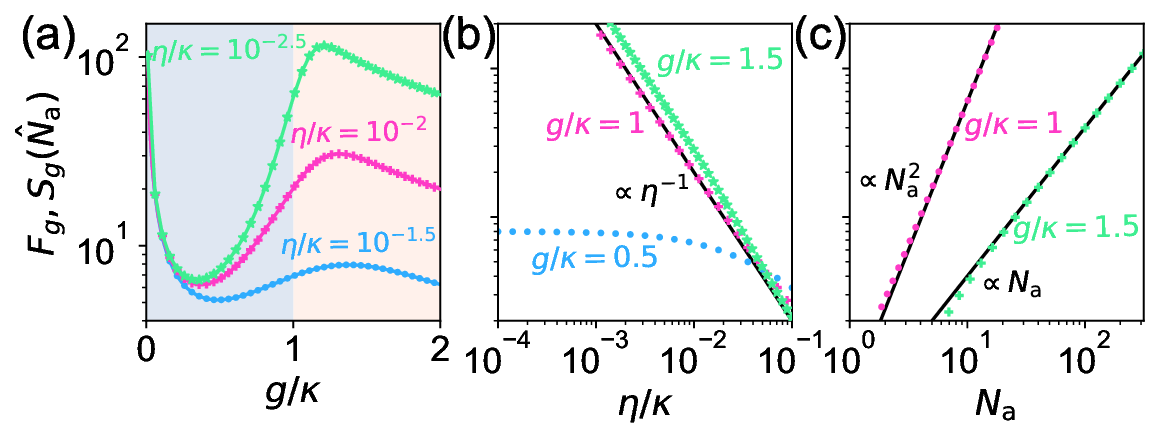}
	\caption{Enhanced metrology. (a) The signal-to-noise ratio with observable $\hat{N}_{\mathrm{a}}$ (dotted lines) and the QFI (solid lines) when estimating the parameter $g$. The pink (blue) area indicates the time crystal (normal) phase. (b) The QFI as a function of $\eta$. The black line is a reference line corresponding to a scaling with exponent $-1$. (c) The QFI as a function of photon number $N_{\mathrm{a}}$. Two black lines are reference lines corresponding to the Heisenberg limit (above) and the standard quantum limit (below).}
	\label{fig:d}
\end{figure}

In Fig. \ref{fig:c}(a), we show the dissipative gap as a function of $g$ for different $\eta$. When $\eta$ approaches zero, a sharp minimum emerges near the critical point and the minimum converges to the critical point gradually. As shown in Fig. \ref{fig:c}(b), the dissipative gaps at the critical point and the minimum both follow a scaling with a critical exponent near $0.5$.

{\it Enhanced metrology.}---Another important finding from the analytical results is the enhanced metrology. In general, for a given observable $\hat{O}$ and an estimated parameter $x$, the estimated precision can be characterized by the signal-to-noise ratio, which is defined as \cite{di_candia_critical_2023}
\begin{equation}
	S_{x}(\hat{O})=\frac{1}{(\Delta x)^{2}}=\frac{|\partial_{x}\langle \hat{O}\rangle|^{2}}{(\Delta \hat{O})^{2}},
\end{equation}
where $\Delta \hat{O}=\sqrt{\langle \hat{O}^{2}\rangle-\langle \hat{O}\rangle^{2}}$ is the standard deviation of the observable $\hat{O}$. The estimated precision is bounded by the QFI \cite{braun_quantum-enhanced_2018,pezze_quantum_2018}. The QFI denotes the maximum precision by choosing the optimal observable. The QFI of an open system can be expressed as a function of the density matrix. As the off-diagonal elements of the steady-state matrix in our model are zero, the QFI can be further simplified as
\begin{equation}
	F_{x}=\sum_{j}\frac{(\partial_{x}p_{j})^{2}}{p_{j}},
\end{equation}
where $p_{j}=\langle j|\rho|j\rangle$ is the diagonal elements of the density matrix.

Fig. \ref{fig:d}(a) shows both the results of signal-to-noise ratio with observable $\hat{N}_{\mathrm{a}}$ (dotted lines) and the QFI (solid lines) when estimating the parameter $g$. There are two surprising findings. First, the signal-to-noise ratio with observable $\hat{N}_{\mathrm{a}}$ exactly matches the QFI. It indicates that $\hat{N}_{\mathrm{a}}$ is the optimal observable with maximum precision for estimated parameter $g$. Second, we find the QFI satisfies the same finite-size scaling $F_{g}\propto\eta^{-1}$ both at the critical point and in the time crystal phase. However, the steady-state photon number has a scaling as $N_{\mathrm{a}}\propto\eta^{-1/2}$ at the critical point [see Eq. (\ref{eq:na-cri2})] and $N_{\mathrm{a}}\propto\eta^{-1}$ in the time crystal phase [see Eq. (\ref{eq:aam})]. It means the metrology sensitivity can approach the Heisenberg limit $F_{g}\propto N_{\mathrm{a}}^{2}$ at the critical point but only the standard quantum limit $F_{g}\propto N_{\mathrm{a}}$ in the time crystal phase [see Fig. \ref{fig:d}(c)]. Similar scaling can also be obtained for the time required to reach the steady state: $F_{g}\propto T^{2}$ at the critical point and $F_{g}\propto T$ in the time crystal phase \cite{sup}.

\begin{table}[b]
	\caption{\label{tab:table1}%
	Scaling exponents of different steady-state variables versus $\eta$. It results in the standard quantum limit $S_{g}(\hat{N}_{\mathrm{a}})\propto N_{a}$ in the time crystal phase and the Heisenberg limit $S_{g}(\hat{N}_{\mathrm{a}})\propto N_{a}^{2}$ at the critical point.}
	\begin{ruledtabular}
		\begin{tabular}{ccccc}
			 &
			$\partial_{g}N_{\mathrm{a}}$ &
			$\Delta \hat{N}_{\mathrm{a}}$ &
			$S_{g}(\hat{N}_{\mathrm{a}})$ &
			$N_{\mathrm{a}}$\\
			\colrule
			critical point & -1 & -1/2 & -1 & -1/2 \\
			time crystal phase & -1 & -1/2 & -1 & -1
		\end{tabular}
	\end{ruledtabular}
\end{table}

As the signal-to-noise ratio with observable $\hat{N}_{\mathrm{a}}$ exactly matches the QFI, we can analyze the limit of the QFI by calculating the analytical limit of the signal-to-noise ratio. First, we can obtain the limit of both the standard deviation and the susceptibility as
\begin{equation}\label{eq:deltaNlimit}
    \lim_{\eta\to0}\Delta{\hat{N}_{\mathrm{a}}}=\begin{cases}
        \frac{\sqrt{g\kappa}}{\kappa-g}, g<\kappa\\
		\sqrt{\frac{\pi-2}{2\pi}\frac{\kappa}{\eta}}, g=\kappa\\
        \sqrt{\frac{g-\kappa}{2\eta}}, g>\kappa
    \end{cases},
\end{equation}
\begin{equation}
    \lim_{\eta\to0}\partial_{g}N_{\mathrm{a}}=\begin{cases}
        \frac{\kappa}{(\kappa-g)^{2}}, g<\kappa\\
		\frac{\pi-2}{2\pi\eta}, g=\kappa\\
        \frac{1}{2\eta}, g>\kappa
    \end{cases}.
\end{equation}
Then the limit of the signal-to-noise ratio is given by \cite{sup}
\begin{equation}\label{eq:lims}
	\lim_{\eta\to0}S_{g}(\hat{N}_{\mathrm{a}})=\begin{cases}
        \frac{\kappa}{g(\kappa-g)^{2}}, g<\kappa\\
		\frac{\pi-2}{2\pi\eta\kappa}, g=\kappa \\
        \frac{1}{2\eta(g-\kappa)}, g>\kappa
    \end{cases}.
\end{equation}
The scaling exponents of these variables versus $\eta$ are shown in Table \ref{tab:table1}. If we consider $\eta$ as a common variable, the scaling exponents of the standard deviation ($\Delta \hat{N}_{\mathrm{a}}$) and susceptibility ($\partial_{g}N_{\mathrm{a}}$) of the photon number are the same at the critical point and for the time crystal phase, resulting in the same scaling exponent of the signal-to-noise ratio. However, the photon number itself exhibits different scaling in two cases. It leads to the enhanced metrology at the critical point. It is more clear when we consider $N_{\mathrm{a}}$ as a variable. We note that $N_{\mathrm{a}}$ changes versus $\eta$ and is not an independent parameter. If we consider $N_\mathrm{a}$ as a common variable, the scaling exponents of the standard deviation ($\Delta \hat{N}_{\mathrm{a}}$) and susceptibility ($\partial_{g}N_{\mathrm{a}}$) of the photon number are double at the critical point compared to the time crystal phase, resulting in an enhancement of sensitivity from standard quantum limit into the Heisenberg limit. The standard deviation and susceptibility of the photon number are both more sensitive to the estimated parameter $g$ at the critical point. However, the enhancements of standard deviation and susceptibility are not counteracted. We obtain a higher scaling exponent of the total estimated sensitivity from larger enhancement of the susceptibility.



{\it Conclusion.}---In conclusion, we present an exact solution for the steady state of the quantum van der Pol oscillator with time crystal phase and dissipative phase transition. From the exact solution, we obtain the detailed critical exponents in the dissipative phase transition. Moreover, we calculate the QFI of the steady state and show the QFI at the critical point and in the time crystal phase both satisfies a scaling $\eta^{-1}$ in the thermodynamic limit. However, due to the different scaling of the steady-state photon number, only the QFI at the critical point can approach the Heisenberg limit $F_{g}\propto N_{\mathrm{a}}^{2}$. The QFI in the time crystal phase can only approach the standard limit $F_{g}\propto N_{\mathrm{a}}$. Similar scaling can also be obtained for the time required to reach the steady state: $F_{g}\propto T^{2}$ at the critical point and $F_{g}\propto T$ in the time crystal phase. We also show the photon number $\hat{N}_{\mathrm{a}}$ is the optimal observable to obtain the maximum sensitivity. It allows us to analyze the origin of the enhanced metrology. In the time crystal phase, the standard-quantum-limit metrology originates from the enhancement of the photon number's susceptibility with a larger scaling than the enhancement of the photon-number standard deviation. At the critical point, the QFI further approaches the Heisenberg limit due to the double scaling of both the standard deviation and susceptibility of the photon number. These results can deepen our understanding of the quantum criticality and universality in open many-body systems.

\begin{acknowledgments}
	This work is supported by the National Key R\&D Program of China (Grant No. 2023YFA1407600), and the National Natural Science Foundation of China (NSFC) (Grants No. 123B2066, No. 12275145, No. 92050110, No. 91736106, No. 11674390, and No. 91836302).
\end{acknowledgments}

%

\appendix

\begin{widetext}

\section{Derivation of the complex $P$-representation of the steady-state density matrix}

We consider a driven single-mode system described by the annihilation operator $a$. The detuning and frequency of the driving field are $\Delta$ and $\varepsilon$. We also assume there is energy pump from one-photon absorption $g$ and damping from both one-photon emission $\kappa$ and two-photon emission $\eta$. The Hamiltonian and master equation can be written as
\begin{gather}
	H=-\Delta \hat{a}^{\dag}\hat{a}+\varepsilon \hat{a}^{\dag}+\varepsilon^{*} \hat{a},\\
	\dot{\rho}=i[\rho,H]+{\kappa}\mathcal{D}(\hat{a})\rho+g\mathcal{D}(\hat{a}^{\dag})\rho+\eta\mathcal{D}(\hat{a}^{2})\rho,
\end{gather}
where $\mathcal{D}(\hat{o})\rho=\hat{o}\rho \hat{o}^{\dag}-(\hat{o}^{\dag}\hat{o}\rho+\rho \hat{o}^{\dag}\hat{o})/2$ is the Lindblad dissipator for operator $\hat{o}$. Afterwards, we consider the complex $P$-representation of the density matrix \cite{drummond_quantum_1980,bartolo_exact_2016}
\begin{equation}\label{eq:p-rho}
	\rho=\int d\alpha\int d\beta \frac{|\alpha \rangle\langle \beta^{*}|}{\langle \beta^{*}|\alpha \rangle}P(\alpha,\beta),
\end{equation}
where the integral is defined in closed contours that encircle all the singularities. To be convenient, we have neglected the integral range. $|\alpha \rangle=e^{-|\alpha|^{2}/2}\sum_{n=0}^{\infty}\frac{\alpha^{n}}{\sqrt{n!}}|n\rangle$ is the coherent state, which satisfies
\begin{gather}
	\hat{a}|\alpha \rangle=\alpha|\alpha \rangle,\langle \beta^{*}|\hat{a}^{\dag}=\beta\langle \beta^{*}|,\\
	\hat{a}^{\dag}|\alpha \rangle=\left(\frac{\alpha^{*}}{2}+\frac{\partial}{\partial\alpha}\right)|\alpha \rangle,\langle \beta^{*}|\hat{a}=\left(\frac{\beta^{*}}{2}+\frac{\partial}{\partial\beta}\right)\langle \beta^{*}|,\\
	\hat{a}^{\dag}\hat{a}|\alpha \rangle=\alpha\left(\frac{\alpha^{*}}{2}+\frac{\partial}{\partial\alpha}\right)|\alpha \rangle,\langle\beta^{*}|\hat{a}^{\dag}\hat{a}=\beta\left(\frac{\beta^{*}}{2}+\frac{\partial}{\partial\beta}\right)\langle \beta^{*}|,\\
	\hat{a}\hat{a}^{\dag}|\alpha \rangle=\left(\frac{\alpha^{*}}{2}+\frac{\partial}{\partial\alpha}\right)\alpha|\alpha \rangle,\langle\beta^{*}|\hat{a}\hat{a}^{\dag}=\left(\frac{\beta^{*}}{2}+\frac{\partial}{\partial\beta}\right)\beta\langle \beta^{*}|.
\end{gather}
Consequently, we can obtain the following replacing rules:
\begin{equation}
	\hat{a}\rho \to \alpha P(\alpha,\beta),\rho \hat{a}^{\dag} \to \beta P(\alpha,\beta),\hat{a}\rho \hat{a}^{\dag} \to \alpha\beta P(\alpha,\beta),\hat{a}^{2}\rho \hat{a}^{\dag2} \to \alpha^{2}\beta^{2} P(\alpha,\beta),
\end{equation}
Because
\begin{equation}
	\begin{split}
		\hat{a}^{\dag}\rho=&\int d\alpha\int d\beta \frac{P(\alpha,\beta)}{\langle \beta^{*}|\alpha \rangle}\left(\frac{\alpha^{*}}{2}+\partial_{\alpha}\right)|\alpha \rangle\langle \beta^{*}|\\
		=&\int d\alpha\int d\beta |\alpha \rangle\langle \beta^{*}|\left(\frac{\alpha^{*}}{2}-\partial_{\alpha}\right)\frac{P(\alpha,\beta)}{\langle \beta^{*}|\alpha \rangle}\\
		=&\int d\alpha\int d\beta |\alpha \rangle\langle \beta^{*}|\frac{P(\alpha,\beta)(\alpha^{*}/2+\partial_{\alpha})\langle \beta^{*}|\alpha \rangle-\langle \beta^{*}|\alpha \rangle\partial_{\alpha}P(\alpha,\beta)}{\langle \beta^{*}|\alpha \rangle^{2}}\\
		=&\int d\alpha\int d\beta |\alpha \rangle\langle \beta^{*}|\frac{(\beta-\partial_{\alpha})P(\alpha,\beta)}{\langle \beta^{*}|\alpha \rangle},
	\end{split}
\end{equation}
we can also obtain
\begin{gather}
	\hat{a}^{\dag}\rho \to (\beta-\partial_{\alpha})P(\alpha,\beta),
\end{gather}
and similarly
\begin{gather}
	\rho \hat{a} \to (\alpha-\partial_{\beta}) P(\alpha,\beta),\\
	\hat{a}\hat{a}^{\dag}\rho \to \alpha(\beta-\partial_{\alpha})P(\alpha,\beta),\rho \hat{a}\hat{a}^{\dag} \to \beta (\alpha-\partial_{\beta}) P(\alpha,\beta),\\
	\hat{a}^{\dag}\hat{a}\rho \to (\beta-\partial_{\alpha})\alpha P(\alpha,\beta),\rho \hat{a}^{\dag}\hat{a} \to (\alpha-\partial_{\beta})\beta P(\alpha,\beta),\\
	\hat{a}^{\dag2}\hat{a}^{2}\rho \to (\beta-\partial_{\alpha})^{2}\alpha^{2} P(\alpha,\beta),\rho \hat{a}^{\dag2}\hat{a}^{2} \to (\alpha-\partial_{\beta})^{2}\beta^{2} P(\alpha,\beta),\\
	\hat{a}^{\dag}\rho \hat{a} \to (\beta-\partial_{\alpha})(\alpha-\partial_{\beta})P(\alpha,\beta).
\end{gather}
Based on the above replacing rules, we can transfer the master equation into a Fokker-Planck equation as
\begin{equation}
	\begin{split}
		i\partial_{t}P& = \partial_{\beta}\left[(-\Delta\beta+\varepsilon^{*})P\right] + \partial_{\alpha}\left[(\Delta\alpha-\varepsilon)P\right] + i\frac{\kappa}{2}\left[\partial_{\alpha}(\alpha P)+\partial_{\beta}(\beta P)\right]\\
		+ i\frac{g}{2}&\left[2\partial_{\alpha}\partial_{\beta}P-\partial_{\alpha}(\alpha P)-\partial_{\beta}(\beta P)\right] + i\frac{\eta}{2}\left\{\partial_{\alpha}\left[(2\beta-\partial_{\alpha})(\alpha^{2}P)\right]+\partial_{\beta}\left[(2\alpha-\partial_{\beta})(\beta^{2} P)\right]\right\}.
	\end{split}
\end{equation}
This equation can also be written as a more concise form as
\begin{equation}
	i\partial_{t}P = \sum_{i=\alpha,\beta}\partial_{i}\left[-A^{i}P+\frac{1}{2}\sum_{j=\alpha,\beta}\partial_{j}(D^{ij}P)\right],
\end{equation}
where
\begin{equation}
	\vec{A} = \begin{pmatrix}
		-\left(\Delta+i\frac{\kappa-g}{2}\right)\alpha-i\eta\alpha^{2}\beta+\varepsilon\\
		\left(\Delta-i\frac{\kappa-g}{2}\right)\beta-i\eta\alpha\beta^{2}-\varepsilon^{*}
	\end{pmatrix},
\end{equation}
\begin{equation}
	\bm{D} = \begin{pmatrix}
		-i\eta\alpha^{2} & ig\\
		ig & -i\eta\beta^{2}
	\end{pmatrix}.
\end{equation}

A possible solution is given by
\begin{equation}
	-A^{i}P+\frac{1}{2}\sum_{j=\alpha,\beta}\partial_{j}(D^{ij}P)=0,
\end{equation}
or
\begin{equation}
	2A^{i}-\sum_{j=\alpha,\beta}\partial_{j}(D^{ij})=\sum_{j=\alpha,\beta}D^{ij}\partial_{j}(\ln P),
\end{equation}
It means
\begin{equation}
	\partial_{j}(\ln P)=\sum_{i=\alpha,\beta}({\bm{D}}^{-1})_{ji}\left[2A^{i}-\sum_{k=\alpha,\beta}\partial_{k}(D^{ik})\right],
\end{equation}
which is equivalent to the following two equations
\begin{gather}
	\partial_{\alpha}(\ln P) = i\left(\frac{-2\Delta \beta}{\eta\alpha\beta+{g}}-i\beta\frac{{\kappa}+g-2\eta}{\eta\alpha\beta-{g}}+2\frac{\eta\beta^{2}\varepsilon-{g}\varepsilon^{*}}{\eta^{2}\alpha^{2}\beta^{2}-{g^{2}}}-2i\beta\right),\\
	\partial_{\beta}(\ln P) = i\left(\frac{2\Delta \alpha}{\eta\alpha\beta+{g}}-i\alpha\frac{{\kappa}+g-2\eta}{\eta\alpha\beta-{g}}-2\frac{\eta\alpha^{2}\varepsilon^{*}-{g}\varepsilon}{\eta^{2}\alpha^{2}\beta^{2}-{g^{2}}}-2i\alpha\right).
\end{gather}
Because $\partial_{\beta}\partial_{\alpha}(\ln P)=\partial_{\alpha}\partial_{\beta}(\ln P)$, the existence of the guessing solution requires
\begin{equation}
	\begin{split}
		&{g}\frac{-2\Delta }{(\eta\alpha\beta+{g})^{2}}+ig\frac{{\kappa}+g-2\eta}{(\eta\alpha\beta-{g})^{2}}+4\eta\beta g\frac{\eta\alpha^{2}\varepsilon^{*}-{g}\varepsilon}{(\eta^{2}\alpha^{2}\beta^{2}-{g^{2}})^{2}}-2i\\
		=&{g}\frac{2\Delta }{(\eta\alpha\beta+{g})^{2}}+ig\frac{{\kappa}+g-2\eta}{(\eta\alpha\beta-{g})^{2}}+4\eta\alpha g\frac{\eta\beta^{2}\varepsilon^{*}-{g}\varepsilon}{(\eta^{2}\alpha^{2}\beta^{2}-{g^{2}})^{2}}-2i.
	\end{split}
\end{equation}

There are two conditions that satisfy the requirement. The first condition is $g=0$, which has been considered in Ref. \cite{drummond_quantum_1980,bartolo_exact_2016}. In this condition, the complex $P$-representation satisfies
\begin{gather}
	\partial_{\alpha}(\ln P) = i\left(\frac{-2\Delta}{\eta\alpha}-2i\frac{\frac{\kappa}{2}-\eta}{\eta\alpha}+\frac{2\varepsilon}{\eta\alpha^{2}}-2i\beta\right),\\
	\partial_{\beta}(\ln P) = i\left(\frac{2\Delta}{\eta\beta}-2i\frac{\frac{\kappa}{2}-\eta}{\eta\beta}-\frac{2\varepsilon^{*}}{\eta\beta^{2}}-2i\alpha\right),
\end{gather}
and the solution is
\begin{equation}
	\ln P(\alpha,\beta) = -(2+d)\ln \alpha - (2+d^{*})\ln \beta + c\alpha^{-1} + c^{*}\beta^{-1}+2\alpha\beta+\mathcal{C},
\end{equation}
or
\begin{equation}
	P(\alpha,\beta) \propto \frac{e^{2\alpha\beta}e^{c/\alpha}e^{c^{*}/\beta}}{\alpha^{2+d}\beta^{2+d^{*}}},
\end{equation}
where $c = -2i\varepsilon/\eta$, $d= 2i(\Delta+i\kappa/2)/\eta$. The normalization coefficient is
\begin{equation}
	\begin{split}
		I=&\int d\alpha\int d\beta P(\alpha,\beta)\\
		=&\int d\alpha\int d\beta \frac{e^{2\alpha\beta}e^{c/\alpha}e^{c^{*}/\beta}}{\alpha^{2+d}\beta^{2+d^{*}}}\\
		=&\int dx \int dy e^{2/(xy)}x^{d}y^{d^{*}}e^{cx+c^{*}y}\\
		=&\int dx \int dy \sum_{k=0}^{\infty}\frac{2^{k}}{k!}x^{d-k}y^{d^{*}-k}e^{cx+c^{*}y}.
	\end{split}
\end{equation}
Using the relation
\begin{equation}
	\left[\Gamma(k-d)\right]^{-1}=\int\frac{(c)^{1+d-k}}{2\pi i}x^{d-k}e^{cx}dx,
\end{equation}
we can simplify the normalization coefficient as
\begin{equation}
	\begin{split}
		I=&-4\pi^{2}\sum_{k=0}^{\infty}\frac{2^{k}(c)^{k-d-1}(c^{*})^{k-d^{*}-1}}{k!\Gamma(k-d)\Gamma(k-d^{*})}\\
		=&\frac{-4\pi^{2}}{(c)^{d+1}(c^{*})^{d^{*}+1}\Gamma(-d)\Gamma(-d^{*})}{_{0}F_{2}}(-d,-d^{*},2|c|^{2}),
	\end{split}
\end{equation}
where $F$ is the generalized hypergeometric function
\begin{equation}
	_{m}F_{n}(a_{1},a_{2},\cdots,a_{m};b_{1},b_{2},\cdots,b_{n},z)=\sum_{k=0}^{\infty}\frac{z^{k}}{k!}\frac{\Gamma(k+a_{1})}{\Gamma(a_{1})}\cdots\frac{\Gamma(k+a_{n})}{\Gamma(a_{n})}\frac{\Gamma(b_{1})}{\Gamma(k+b_{1})}\cdots\frac{\Gamma(b_{n})}{\Gamma(k+b_{n})}.
\end{equation}
The Wigner equation is then given by
\begin{equation}
	\begin{split}
		W(z)&=\frac{2}{\pi}\int d\alpha\int d\beta P(\alpha,\beta)e^{-2(\alpha-z)(\beta-z^{*})}\\
		&=\frac{2}{\pi I}e^{-2|z|^{2}}\int d\alpha\int d\beta \frac{e^{c/\alpha}e^{c^{*}/\beta}}{\alpha^{2+d}\beta^{2+d^{*}}}e^{-2\alpha z^{*}}e^{-2\beta z}\\
		&=\frac{8\pi}{I}e^{-2|z|^{2}}\left|\sum_{k=0}^{\infty}\frac{(2z^{*})^{k}(c)^{k-d-1}}{k!\Gamma(k-d)}\right|^{2}\mathrm{or}=\frac{8\pi}{I}e^{-2|z|^{2}}\left|\sum_{k=0}^{\infty}\frac{(2z^{*})^{k+d+1}(c)^{k}}{k!\Gamma(k+d+2)}\right|^{2}\\
		&=\frac{8\pi}{I}e^{-2|z|^{2}}\left|\frac{_{0}F_{1}(-d,2cz^{*})}{(c)^{d+1}\Gamma(-d)}\right|^{2}\mathrm{or}=\frac{8\pi}{I}e^{-2|z|^{2}}\left|(2z^{*})^{d+1}\frac{_{0}F_{1}(d+2,2cz^{*})}{\Gamma(d+2)}\right|^{2}.
	\end{split}
\end{equation}

The second condition is what we considered: $g\neq0$ but $\Delta=\varepsilon=0$. It means
\begin{gather}
	\partial_{\alpha}(\ln P) = i\left(-i\beta\frac{{\kappa}+g-2\eta}{\eta\alpha\beta-{g}}-2i\beta\right),\\
	\partial_{\beta}(\ln P) = i\left(-i\alpha\frac{{\kappa}+g-2\eta}{\eta\alpha\beta-{g}}-2i\alpha\right).
\end{gather}
The solution is
\begin{equation}
	\ln P(\alpha,\beta) = \frac{\kappa+g-2\eta}{\eta}\ln(\eta\alpha\beta-{g})+2\alpha\beta+\mathcal{C},
\end{equation}
or
\begin{equation}
	P(\alpha,\beta) \propto e^{2\alpha\beta}(\eta\alpha\beta-{g})^{\frac{\kappa+g-2\eta}{\eta}}.
\end{equation}
The normalization coefficient is (letting $\frac{\kappa+g-2\eta}{\eta}\equiv q$)
\begin{equation}
	\begin{split}
		I=&\int d\alpha\int d\beta P(\alpha,\beta)\\
		=&\int d\alpha\int d\beta e^{2\alpha\beta}(\eta\alpha\beta-{g})^{q}\\
		=&\int d\alpha\int d\beta \sum_{k=0}^{\infty}\frac{2^{k}}{k!}(\eta\alpha\beta-{g})^{q}\alpha^{k}\beta^{k}\\
		=&\sum_{k=0}^{\infty}\frac{2^{k}}{k!}\mathcal{F}_{k,k}(\eta,g,q),
	\end{split}
\end{equation}
where
\begin{equation}
	\begin{split}
		\mathcal{F}_{m,n}(\eta,g,q)=&\int d\alpha\int d\beta (\eta\alpha\beta-{g})^{q}\alpha^{m}\beta^{n}\\
        =&\int d\gamma \int d\beta (-g)^{q}\left(\frac{g}{\eta}\right)^{m+1}(1-\gamma)^{q}\gamma^{m}\beta^{n-m-1}\\
        =&2\pi i(-g)^{q}\left(\frac{g}{\eta}\right)^{m+1}\delta_{mn}\int d\gamma (1-\gamma)^{q}\gamma^{m}\\
        =&2\pi i(-g)^{q}\left(\frac{g}{\eta}\right)^{m+1}\delta_{mn}\frac{\Gamma(q+1)\Gamma(m+1)}{\Gamma(m+q+1)}\int d\gamma (1-\gamma)^{q+m}\\
		=&2\pi i\left(-{g}\right)^{q}\left(\frac{g}{\eta}\right)^{m+1}\delta_{mn}\frac{\Gamma(q+1)\Gamma(m+1)}{\Gamma(m+q+1)}\frac{1-e^{2\pi iq}}{m+q+1}\\
		=&\mathcal{N}\left(\frac{g}{\eta}\right)^{m}\frac{\Gamma(q+2)\Gamma(m+1)}{\Gamma(m+q+2)}\delta_{mn}.
	\end{split}
\end{equation}
$\mathcal{N}$ is a constant that is independent to $m$ and $n$. Then we can simplify the normalization coefficient as
\begin{equation}\label{eq:norm}
	\frac{I}{\mathcal{N}}={_{1}F_{1}}(1;q+2,\frac{2g}{\eta}).
\end{equation}
The Wigner equation is given by
\begin{equation}
	\begin{split}
		W(z)=&\frac{2}{\pi}\int d\alpha\int d\beta P(\alpha,\beta)e^{-2(\alpha-z)(\beta-z^{*})}\\
		=&\frac{2}{\pi I}e^{-2|z|^{2}}\int d\alpha\int d\beta (\eta\alpha\beta-{g})^{q}e^{2\alpha z^{*}}e^{2\beta z}\\
		=&\frac{2}{\pi I}e^{-2|z|^{2}}\int d\alpha\int d\beta \sum_{m,n=0}^{\infty}(\eta\alpha\beta-{g})^{q}\frac{(2\alpha z^{*})^{m}(2\beta z)^{n}}{m!n!}\\
		=&\frac{2}{\pi I}e^{-2|z|^{2}}\sum_{m,n=0}^{\infty}\frac{(2 z^{*})^{m}(2 z)^{n}}{m!n!}\mathcal{F}_{m,n}(\eta,g,q)\\
		=&\frac{2}{\pi I}e^{-2|z|^{2}}\sum_{m,n=0}^{\infty}\frac{(2 z^{*})^{m}(2 z)^{n}}{m!n!}\mathcal{N}\left(\frac{g}{\eta}\right)^{m}\frac{\Gamma(q+2)\Gamma(m+1)}{\Gamma(m+q+2)}\delta_{mn}\\
		=&\frac{2\mathcal{N}}{\pi I}e^{-2|z|^{2}}{_{0}F_{1}}(q+2,\frac{4g|z|^{2}}{\eta}),
	\end{split}
\end{equation}
and the factorial moments is
\begin{equation}\label{eq:fm}
	\begin{split}
		\langle \hat{a}^{\dag m}\hat{a}^{n}\rangle=&\int d\alpha\int d\beta P(\alpha,\beta)\alpha^{n}\beta^{m}\\
		=&\frac{1}{I}\int d\alpha\int d\beta e^{2\alpha\beta}(\eta\alpha\beta-{g})^{q}\alpha^{n}\beta^{m}\\
		=&\frac{1}{I}\int d\alpha\int d\beta \sum_{k=0}^{\infty}\frac{2^{k}}{k!}(\eta\alpha\beta-{g})^{q}\alpha^{k+n}\beta^{k+m}\\
		=&\frac{1}{I}\sum_{k=0}^{\infty}\frac{2^{k}}{k!}\mathcal{F}_{k+n,k+m}(\eta,g,q)\\
		=&\frac{1}{I}\sum_{k=0}^{\infty}\frac{2^{k}}{k!}\mathcal{N}\left(\frac{g}{\eta}\right)^{k+m}\frac{\Gamma(q+2)\Gamma(k+m+1)}{\Gamma(k+m+q+2)}\delta_{mn}\\
		=&\frac{\mathcal{N}}{I}\left(\frac{g}{\eta}\right)^{m}\frac{\Gamma(q+2)\Gamma(m+1)}{\Gamma(m+q+2)}{_{1}F_{1}}(m+1;m+q+2,\frac{2g}{\eta})\delta_{mn}.
	\end{split}
\end{equation}
The density matrix is
\begin{equation}
	\begin{split}
		\rho=&\int d\alpha\int d\beta \frac{|\alpha \rangle\langle \beta^{*}|}{\langle \beta^{*}|\alpha \rangle}P(\alpha,\beta)\\
		=&\frac{1}{I}\int d\alpha\int d\beta \frac{|\alpha \rangle\langle \beta^{*}|}{\langle \beta^{*}|\alpha \rangle}e^{2\alpha\beta}(\eta\alpha\beta-{g})^{q}\\
		=&\frac{1}{I}\int d\alpha\int d\beta e^{\alpha\beta}\sum_{n,m}\frac{\alpha^{n}\beta^{m}}{\sqrt{n!m!}}|n\rangle\langle m|(\eta\alpha\beta-{g})^{q}\\
		=&\frac{1}{I}\sum_{n,m}\frac{1}{\sqrt{n!m!}}|n\rangle\langle m|\int d\alpha\int d\beta e^{\alpha\beta}\alpha^{n}\beta^{m}(\eta\alpha\beta-{g})^{q}\\
		=&\frac{1}{I}\sum_{n,m}\frac{1}{\sqrt{n!m!}}|n\rangle\langle m|\int d\alpha\int d\beta \sum_{k=0}^{\infty}\frac{1}{k!}\alpha^{k+n}\beta^{k+m}(\eta\alpha\beta-{g})^{q}\\
		=&\frac{1}{I}\sum_{n,m}\frac{1}{\sqrt{n!m!}}|n\rangle\langle m|\sum_{k=0}^{\infty}\frac{1}{k!}\mathcal{F}_{k+n,k+m}(\eta,g,q)\\
		=&\frac{1}{I}\sum_{n,m}\frac{1}{\sqrt{n!m!}}|n\rangle\langle m|\sum_{k=0}^{\infty}\frac{1}{k!}\mathcal{N}\left(\frac{g}{\eta}\right)^{k+m}\frac{\Gamma(q+2)\Gamma(k+m+1)}{\Gamma(k+m+q+2)}\delta_{mn}\\
		=&\sum_{n,m}\frac{1}{\sqrt{n!m!}}|n\rangle\langle m|\frac{\mathcal{N}}{I}\left(\frac{g}{\eta}\right)^{m}\frac{\Gamma(q+2)\Gamma(m+1)}{\Gamma(m+q+2)}{_{1}F_{1}}(m+1;m+q+2,\frac{g}{\eta})\delta_{mn}\\
		=&\sum_{n,m}|n\rangle\langle m|\frac{\mathcal{N}}{I}\left(\frac{g}{\eta}\right)^{m}\frac{\Gamma(q+2)}{\Gamma(m+q+2)}{_{1}F_{1}}(m+1;m+q+2,\frac{g}{\eta})\delta_{mn}.
	\end{split}
\end{equation}

\section{Derivation of the limit for $\eta\to0$ and $g\neq\kappa$}

Firstly, we derive the limit of the factorial moments, then the other limit can be obtained straightforwardly. The factorial moments can be obtained as [substituting Eq. (\ref{eq:norm}) into Eq. (\ref{eq:fm})]
\begin{equation}\label{eq:nmfp}
    \begin{split}
        \langle \hat{a}^{\dag m}\hat{a}^{m}\rangle=&\frac{m!\left(\frac{g}{\eta}\right)^{m}{_{1}F_{1}}\left(1+m;\frac{\kappa+g}{\eta}+m,\frac{2g}{\eta}\right)}{\left(\frac{\kappa+g}{\eta}\right)_{m}{_{1}F_{1}}\left(1;\frac{\kappa+g}{\eta},\frac{2g}{\eta}\right)},
    \end{split}
\end{equation}
where $(x)_{n}=x(x+1)\cdots(x+n-1)$ is the Pochhammer symbol. The above equation can be divided into two parts: the generalized hypergeometric function and the coefficient. The limit of the coefficient can be easily obtained as
\begin{equation}\label{eq:limits}
	\lim_{\eta\to0}\frac{m!\left(\frac{g}{\eta}\right)^{m}}{\left(\frac{\kappa+g}{\eta}\right)_{m}}=m!\left(\frac{g}{\kappa+g}\right)^{m}.
\end{equation}
The limit of the generalized hypergeometric function should be derived under two conditions respectively: $g<\kappa$ and $g>\kappa$. We first write down the summation form of the generalized hypergeometric function.
\begin{equation}
	{_{1}F_{1}}\left(1+m;\frac{\kappa+g}{\eta}+m,\frac{2g}{\eta}\right)=\sum_{k=0}^{\infty}\frac{1}{k!}\frac{(1+m)_{k}}{\left(\frac{\kappa+g}{\eta}+m\right)_{k}}\left(\frac{2g}{\eta}\right)^{k}.
\end{equation}
The index $k_{0}$ of the maximum value in the summation satisfies
\begin{equation}
	(1+m+k_{0})\frac{2g}{\eta}=\left(\frac{\kappa+g}{\eta}+m+k_{0}\right)k_{0}.
\end{equation}
In the limit $\eta\to0$, we can approximately obtain
\begin{equation}
    \lim_{\eta\to0}k_{0}\approx\begin{cases}
        \frac{2g}{\kappa-g}(1+m)\ll\frac{\kappa}{\eta}, g<\kappa\\
        \frac{g-\kappa}{\eta}\sim\frac{\kappa}{\eta}, g>\kappa
    \end{cases}.
\end{equation}
It means that for $g<\kappa$ we only need to consider a few number of series in the summation compared with $\frac{\kappa}{\eta}$. So 
\begin{equation}\label{eq:limitf1}
	\begin{split}
		&\lim_{\eta\to0}{_{1}F_{1}}\left(1+m;\frac{\kappa+g}{\eta}+m,\frac{2g}{\eta}\right)\\
		=&\lim_{\eta\to0}\sum_{k=0}^{\infty}\frac{1}{k!}\frac{(1+m)_{k}}{\left(\frac{\kappa+g}{\eta}+m\right)_{k}}\left(\frac{2g}{\eta}\right)^{k}\\
		=&\sum_{k=0}^{\infty}\frac{(1+m)_{k}}{k!}\left(\frac{2g}{\kappa+g}\right)^{k}\\
		=&\left(1-\frac{2g}{\kappa+g}\right)^{-1-m}.
	\end{split}
\end{equation}
Substituting Eq. (\ref{eq:limits}) and Eq. (\ref{eq:limitf1}) into Eq. (\ref{eq:nmfp}), we obtain the limit for $g<\kappa$:
\begin{equation}
	\lim_{\eta\to0}\langle \hat{a}^{\dag m}\hat{a}^{m}\rangle=\frac{m!g^{m}}{(\kappa-g)^{m}}.
\end{equation}

In the condition $g>\kappa$, the summation is unsolvable. However, we can still derive the limit through the integral representation of the hypergeometric function, which is 
\begin{equation}\label{eq:integral}
	{_{1}F_{1}}\left(1+m;\frac{\kappa+g}{\eta}+m,\frac{2g}{\eta}\right)=\frac{\Gamma(\frac{\kappa+g}{\eta}+m)}{\Gamma(1+m)\Gamma(\frac{\kappa+g}{\eta}-1)}\int_{0}^{1}e^{\frac{2gt}{\eta}}t^{m}(1-t)^{\frac{\kappa+g}{\eta}-2}dt.
\end{equation}
The maximum point $t_{0}$ in the integral function satisfies
\begin{equation}
	\frac{2g}{\eta}+\frac{m}{t}+\frac{\frac{\kappa+g}{\eta}-2}{t-1}=0.
\end{equation}
For small $\eta$, the maximum point is
\begin{equation}
	t_{0}=\frac{g-\kappa}{2g}+{o}(1),
\end{equation}
where ${o}(\eta)$ represents the higher-order series that can be neglected in the limit $\eta\to0$. According to the Laplace method, the integral can be expressed as
\begin{equation}
	\int_{0}^{1}e^{\frac{2gt}{\eta}}t^{m}(1-t)^{\frac{\kappa+g}{\eta}-2}dt= e^{S(t_{0})}\left(\sqrt{\frac{2\pi}{-S^{\prime\prime}(t_{0})}}+{o}(\eta^{1/2})\right),
\end{equation}
where
\begin{equation}
	S(t)=\log\left[e^{\frac{2gt}{\eta}}t^{m}(1-t)^{\frac{\kappa+g}{\eta}-2}\right].
\end{equation}
Moreover, because
\begin{equation}
	\lim_{\eta\to0}\frac{\Gamma(\frac{\kappa+g}{\eta}+m)}{\Gamma(\frac{\kappa+g}{\eta}-1)}=\left(\frac{\kappa+g}{\eta}\right)^{1+m},
\end{equation}
the limit of the hypergeometric function is
\begin{equation}\label{eq:limitf2}
	\lim_{\eta\to0}{_{1}F_{1}}\left(1+m;\frac{\kappa+g}{\eta}+m,\frac{2g}{\eta}\right)=\frac{1}{m!}\left(\frac{g^{2}-\kappa^{2}}{2\eta g}\right)^{m}e^{\frac{g-\kappa}{\eta}}\left(\frac{\kappa+g}{2g}\right)^{\frac{\kappa+g}{\eta}-1}\sqrt{2\pi\left(\frac{\kappa+g}{\eta}\right)}.
\end{equation}
Substituting Eq. (\ref{eq:limits}) and Eq. (\ref{eq:limitf2}) into Eq. (\ref{eq:nmfp}), we obtain the limit for $g>\kappa$:
\begin{equation}
	\lim_{\eta\to0}\langle \hat{a}^{\dag m}\hat{a}^{m}\rangle=\left(\frac{g-\kappa}{2\eta}\right)^{m}.
\end{equation}

In conclusion, the limit of the factorial moments are
\begin{equation}
    \lim_{\eta\to0}\langle \hat{a}^{\dag m}\hat{a}^{m}\rangle=\begin{cases}
        \frac{\Gamma(m+1)g^{m}}{(\kappa-g)^{m}}, g<\kappa\\
        \left(\frac{g-\kappa}{2\eta}\right)^{m}, g>\kappa
    \end{cases}.
\end{equation}
Thus, the limit of the photon number is
\begin{equation}\label{eq:photonlimit}
    \lim_{\eta\to0}\langle \hat{a}^{\dag }\hat{a}\rangle=\begin{cases}
        \frac{g}{\kappa-g}, g<\kappa\\
        \frac{g-\kappa}{2\eta}, g>\kappa
    \end{cases},
\end{equation}
and the limit of the susceptibility of the photon number is
\begin{equation}
    \lim_{\eta\to0}\partial_{g}\langle \hat{a}^{\dag }\hat{a}\rangle=\begin{cases}
        \frac{\kappa}{(\kappa-g)^{2}}, g<\kappa\\
        \frac{1}{2\eta}, g>\kappa
    \end{cases},
\end{equation}
For $m=2$, the limit of the factorial moment is
\begin{equation}
    \lim_{\eta\to0}\langle \hat{a}^{\dag 2}\hat{a}^{2}\rangle=\begin{cases}
        \frac{2g^{2}}{(\kappa-g)^{2}}, g<\kappa\\
        \frac{(g-\kappa)^{2}}{4\eta^{2}}, g>\kappa
    \end{cases}.
\end{equation}
So the limit of the photon number fluctuation $\Delta{\hat{N}_{\mathrm{a}}}=\sqrt{\langle \hat{a}^{\dag 2}\hat{a}^{2}\rangle+\langle \hat{a}^{\dag }\hat{a}\rangle-\langle \hat{a}^{\dag }\hat{a}\rangle^{2}}$ is
\begin{equation}\label{eq:deltaNlimit}
    \lim_{\eta\to0}\Delta{\hat{N}_{\mathrm{a}}}=\begin{cases}
        \frac{\sqrt{g\kappa}}{\kappa-g}, g<\kappa\\
        \sqrt{\frac{g-\kappa}{2\eta}}, g>\kappa
    \end{cases},
\end{equation}
and the limit of the second-order correlation function $g^{(2)}=\langle \hat{a}^{\dag 2}\hat{a}^{2}\rangle/\langle \hat{a}^{\dag }\hat{a}\rangle^{2}$ is
\begin{equation}\label{eq:gtwolimit}
    \lim_{\eta\to0}g^{(2)}=\begin{cases}
        2, g<\kappa\\
        1, g>\kappa
    \end{cases}.
\end{equation}
Then the signal-to-noise ratio with observable $\hat{N}_{\mathrm{a}}$ is given by
\begin{equation}\label{eq:lims}
	\begin{split}
		\lim_{\eta\to0}S_{g}(\hat{N}_{\mathrm{a}})=&\lim_{\eta\to0}\frac{|\partial_{g}\langle \hat{N}_{\mathrm{a}}\rangle|^{2}}{(\Delta \hat{N}_{\mathrm{a}})^{2}}\\
		=&\begin{cases}
			\frac{\kappa}{g(\kappa-g)^{2}}, g<\kappa\\
			\frac{1}{2\eta(g-\kappa)}, g>\kappa
		\end{cases}.
	\end{split}
\end{equation}
Furthermore, we can also obtain the limit of the density matrix for $g<\kappa$ as
\begin{equation}
    \lim_{\eta\to0}p_{n}= \frac{(\kappa-g)g^{n}}{\kappa^{n+1}}.
\end{equation}
The method is similar. It is an exponential distribution, and the large term appears for small $n$. However, when $g>\kappa$, the maximum entry appears when $n=\frac{g-\kappa}{2\eta}$ (obtained by numerical results). It makes the derivation of the limit in this case very difficult. On the other hand, according to the second-order correlation function, the diagonal entries of the density matrix should satisfy the Poisson distribution, which is
\begin{equation}
    \lim_{\eta\to0}p_{n}\approx\frac{1}{n!}\left(\frac{g-\kappa}{2\eta}\right)^{n}e^{-\frac{g-\kappa}{2\eta}}.
\end{equation}
Here we use $\approx$ because this result is not obtained analytically.

\section{Derivation of the finite-size behavior for $g=\kappa$}

When $g=\kappa$, we can also derive the limit through the integral representation Eq. (\ref{eq:integral}). But the Laplace method is not enough here. The derivation is as follows:
\begin{equation}
	\begin{split}
		& \int_{0}^{1}e^{\frac{2\kappa t}{\eta}}t^{m}(1-t)^{\frac{2\kappa}{\eta}-2}dt \\
		=& \int_{0}^{1}t^{m}\exp\left[\frac{2\kappa t}{\eta}+\left(\frac{2\kappa}{\eta}-2\right)\ln(1-t)\right]dt\\
		=& \int_{0}^{1}t^{m}\exp\left[\frac{2\kappa t}{\eta}+\left(\frac{2\kappa}{\eta}-2\right)\left(-t-\frac{t^{2}}{2}+o(t^{2})\right)\right]dt \\
		=& \int_{0}^{1}t^{m}\exp\left[-\frac{\kappa}{\eta}t^{2}+o\left(\frac{t^{2}}{\eta}\right)\right]dt \\
		=& \left[\frac{1}{2}\left(\frac{\kappa}{\eta}\right)^{-(1+m)/2}+o(\eta^{-(1+m)/2})\right]\int_{0}^{\infty}x^{(m-1)/2}e^{-x}dx \\
		=& \Gamma\left(\frac{m+1}{2}\right)\left[\frac{1}{2}\left(\frac{\kappa}{\eta}\right)^{-(1+m)/2}+o(\eta^{(1+m)/2})\right].
	\end{split}
\end{equation}
So, the limit of the hypergeometric function is
\begin{equation}\label{eq:limitf3}
	\lim_{\eta\to0}{_{1}F_{1}}\left(1+m;\frac{\kappa+g}{\eta}+m,\frac{2g}{\eta}\right)=\frac{2^{m}}{m!}\Gamma\left(\frac{m+1}{2}\right)\left(\frac{\kappa}{\eta}\right)^{(1+m)/2}.
\end{equation}
Substituting Eq. (\ref{eq:limits}) and Eq. (\ref{eq:limitf3}) into Eq. (\ref{eq:nmfp}), we obtain the limit for $g=\kappa$:
\begin{equation}
	\lim_{\eta\to0}\langle \hat{a}^{\dag m}\hat{a}^{m}\rangle=\frac{\Gamma\left[(m+1)/2\right]}{\Gamma\left(1/2\right)}\left(\frac{\kappa}{\eta}\right)^{m/2}.
\end{equation}

Thus, the limit of the photon number is
\begin{equation}
    \lim_{\eta\to0}\langle \hat{a}^{\dag }\hat{a}\rangle=\sqrt{\frac{\kappa}{\pi\eta}}.
\end{equation}
For $m=2$, the limit of the factorial moment is
\begin{equation}
    \lim_{\eta\to0}\langle \hat{a}^{\dag 2}\hat{a}^{2}\rangle=\frac{\kappa}{2\eta}.
\end{equation}
So the limit of the photon number fluctuation $\Delta{\hat{N}_{\mathrm{a}}}=\sqrt{\langle \hat{a}^{\dag 2}\hat{a}^{2}\rangle+\langle \hat{a}^{\dag }\hat{a}\rangle-\langle \hat{a}^{\dag }\hat{a}\rangle^{2}}$ is
\begin{equation}
    \lim_{\eta\to0}\Delta{\hat{N}_{\mathrm{a}}}=\sqrt{\left(\frac{1}{2}-\frac{1}{\pi}\right)\frac{\kappa}{\eta}},
\end{equation}
and the limit of the second-order correlation function $g^{(2)}=\langle \hat{a}^{\dag 2}\hat{a}^{2}\rangle/\langle \hat{a}^{\dag }\hat{a}\rangle^{2}$ is
\begin{equation}
    \lim_{\eta\to0}g^{(2)}=\frac{\pi}{2}.
\end{equation}
To derive the finite-size behavior of the signal-to-noise ratio, we need to derive the finite-size behavior of the derivatives of the factorial moments. The dominated term is the derivatives of the integral:
\begin{equation}
	\begin{split}
		& \frac{\partial}{\partial g}\int_{0}^{1}e^{\frac{2gt}{\eta}}t^{m}(1-t)^{\frac{\kappa+g}{\eta}-2}dt \\
		=& \frac{2}{\eta}\int_{0}^{1}e^{\frac{2gt}{\eta}}t^{m+1}(1-t)^{\frac{\kappa+g}{\eta}-2}dt+\frac{1}{\eta}\int_{0}^{1}\ln(1-t)e^{\frac{2gt}{\eta}}t^{m}(1-t)^{\frac{\kappa+g}{\eta}-2}dt \\
		=& \frac{1}{\eta}\int_{0}^{1}e^{\frac{2gt}{\eta}}\left[t^{m+1}+o(t^{m+1})\right](1-t)^{\frac{\kappa+g}{\eta}-2}dt \\
		=& \frac{1}{\eta}\Gamma\left(\frac{m}{2}+1\right)\left[\frac{1}{2}\left(\frac{\kappa}{\eta}\right)^{-m/2-1}+o\left(\eta^{m/2+1}\right)\right].
	\end{split}
\end{equation}
So the derivative of the hypergeometric function is
\begin{equation}
	\lim_{\eta\to0}\frac{\partial}{\partial g}{_{1}F_{1}}\left(1+m;\frac{\kappa+g}{\eta}+m,\frac{2g}{\eta}\right)=\frac{2^{m}}{m!\eta}\left(\frac{\kappa}{\eta}\right)^{m/2}\Gamma\left(\frac{m}{2}+1\right).
\end{equation}
So the derivatives of the factorial moments are
\begin{equation}
	\lim_{\eta\to0}\frac{\partial}{\partial g}\langle \hat{a}^{\dag m}\hat{a}^{m}\rangle=\frac{1}{\eta}\left(\frac{\kappa}{\eta}\right)^{(m-1)/2}\frac{\Gamma(1/2)\Gamma(m/2+1)-\Gamma[(m+1)/2]}{\Gamma(1/2)^{2}}.
\end{equation}
For $m=1$, the derivative of the photon number is
\begin{equation}
	\lim_{\eta\to0}\frac{\partial}{\partial g}N_{\mathrm{a}}=\frac{\pi-2}{2\pi\eta},
\end{equation}
and we can obtain the limit of the signal-to-noise ratio as
\begin{equation}
	\lim_{\eta\to0}S_{g}(\hat{N}_{\mathrm{a}})=\frac{\pi-2}{2\pi\eta\kappa}.
\end{equation}

\section{Discussion about the time required to reach the steady state}

We obtain divergent quantum Fisher information in the thermodynamic limit both at the critical point and in the time crystal phase. However, the enhancement of measurement sensitivity is accompanied by the increasing of time used, as known as the critical slowing down \cite{garbe_critical_2020}. The increasing of time originates from the closing of Liouvillian gap. In Fig. 3 of the main text, we show the closing of real dissipative gap $\Delta_{\mathrm{RDG}}$ at the critical point, and $\Delta_{\mathrm{RDG}}$ follows a half scaling versus $\eta$. The closing of real dissipative gap corresponds to the degeneracy of steady state and the emergence of dissipative phase transition. However, to characterize the time required to reach the steady state, we should consider the asymptotic decay rate ($\Delta_{\mathrm{ADR}}$) \cite{kessler_dissipative_2012}, which is defined as the opposite of the largest real part of all the nonzero eigenvalues. As shown in Fig. \ref{fig:a}, the asymptotic decay rate approaches zero both at the critical point ($g=\kappa$) and in the time crystal phase ($g>\kappa$). However, the asymptotic decay rate exhibits different scaling behaviors in the two cases, $\Delta_{\mathrm{ADR}}\propto \eta^{0.5}$ for $g=\kappa$ and $\Delta_{\mathrm{ADR}}\propto \eta$ for $g>\kappa$. It results in that the time required to reach the steady state also exhibits different scaling behaviors, $T\propto \eta^{-0.5}$ for $g=\kappa$ and $T\propto \eta^{-1}$ for $g>\kappa$, as shown in Table \ref{tab:table1}. Consequently, the quantum Fisher information satisfies the same scaling as a function of both $N_{\mathrm{a}}$ and $T$. The quantum Fisher information approaches the Heisenberg limit at the critical point ($g=\kappa$) and approaches the standard quantum limit in the time crystal phase ($g>\kappa$), both for
$N_{\mathrm{a}}$ and $T$.

Here we can only provide analytical results of the scaling analysis when considering $N_{\mathrm{a}}$. The scaling behavior for $T$ is obtained by numerical fitting. A possible analytical way is using the formalism of third quantization and employing some approximation, which may be finished in subsequent research.

\begin{figure}[h]
	\centering
	\includegraphics[width=0.6\textwidth]{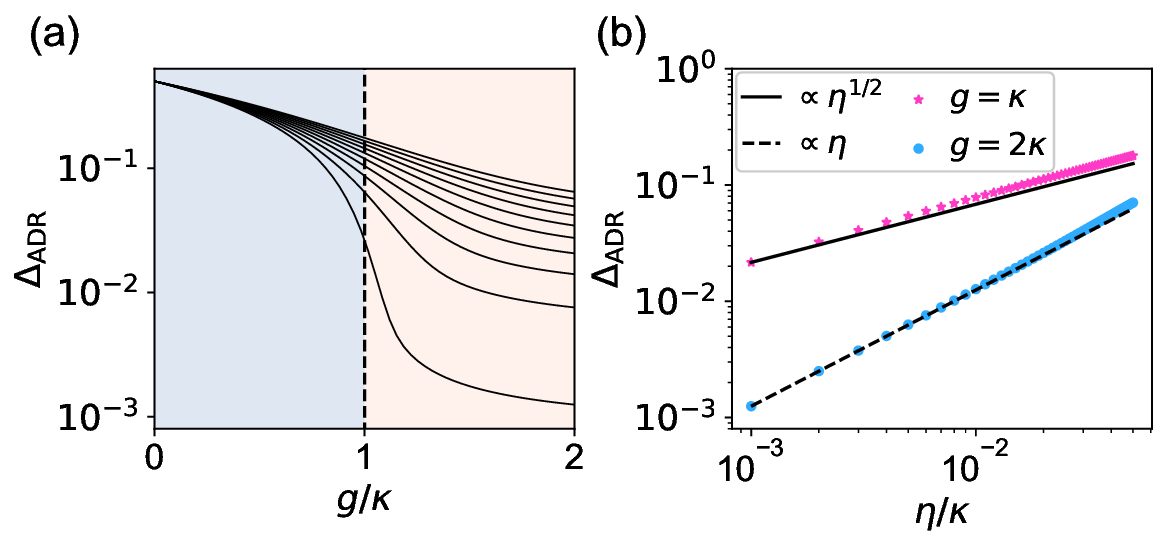}
	\caption{(a) Asymptotic decay rate as a function of $g$ for different  $\eta$. The black dashed line indicates the critical point, and the pink (blue) area indicates the time crystal (normal) phase. The ten lines from the bottom up correspond to equally spaced $\eta/\kappa$ from 0.001 to 0.05. (b) Asymptotic decay rate for $g=\kappa$ (red dotted line) and $g=2\kappa$ (blue dotted line). The black solid (dashed) line is a reference line corresponding to the critical exponent $1/2$ $(1)$.}
	\label{fig:a}
\end{figure}

\begin{table}[h]
	\caption{\label{tab:table1}%
	Scaling exponents of different steady-state variables versus $\eta$. It results in the quantum standard limit $S_{g}(\hat{N}_{\mathrm{a}})\propto N_{a} [T]$ in the time crystal phase and the Heisenberg limit $S_{g}(\hat{N}_{\mathrm{a}})\propto N_{a}^{2} [T^{2}]$ at the critical point.}
	\begin{ruledtabular}
		\begin{tabular}{ccccc}
			 &
			$S_{g}(\hat{N}_{\mathrm{a}})$ &
			$N_{\mathrm{a}}$ &
			$T$\\
			\colrule
			critical point & -1 & -1/2 & -0.5\\
			time crystal phase & -1 & -1 & -1
		\end{tabular}
	\end{ruledtabular}
\end{table}

\end{widetext}

\end{document}